\documentstyle[psfig,epsf]{article}
\begin{document}

\def\Journal#1#2#3#4{{#1} {\bf #2} (#3) #4}
\def \PRL      {Phys. Rev. Lett.~}
\def \PR       {Phys. Rev.}
\def \PRD      {Phys. Rev. D}
\def \PL       {Phys. Lett.~}
\def \PLB      {Phys. Lett. B}
\def \ZPC      {Z. Phys. C}     
\def \NPB      {Nucl. Phys. B}
\def \NPD      {Nucl. Phys. D}
\def \PR       {Phys. Rep.~}
\def \INC      {Il Nuovo Cimento}
\def \NIM      {Nucl. Instrum. Methods}
\def \NIMA     {Nucl. Instrum. Methods Phys. Res. Sect. A}
\def \CPC      {Comput. Phys. Commun.}
\def \EUR      {Eur. Phys. J. C}
\def \PTP      {Prog. Theor. Phys.}
\def \PTPS     {Prog. Theor. Phys. Suppl.}
\def \JHEP     {J. High Energy Phys.}
\def \etal     {\relax\ifmmode{et \; al.}\else{$et \; al.$}\fi}

\begin{center}
{\Large \bf Introduction to GR@PPA event generators for $pp$/$p\bar{p}$ 
collisions} \\

\vspace{4mm}

Soushi Tsuno \\
Department of Physics, Faculty of Science, Okayama University \\ 
3-1-1 Tsushima-naka, Okayama 700-8530, Japan \\
\end{center}

\begin{abstract}
We have developed an extended framework, named GR@PPA, of the GRACE system for 
hadron collisions. The GRACE system is an automatic Feynman diagram 
calculation system and an event generator based on this diagram calculation. 
While the original GRACE system assumes that both the initial and final states 
are well-defined, the GR@PPA framework applies that the initial and final 
states parton configuration is treated in the Feynman diagram calculation at 
the same time by putting one more integration variables. As a result, some 
subprocesses with the same coupling order in hadron-hadron collisions can 
share an identical "GRACE output code" and can be treated as a single 
subprocess. This technique simplifies the program code and saves the computing 
time very much. The constructed event generators would be suitable for the 
large scale Monte Carlo production in the hadron colliders. In this paper, we 
discuss this technique, and present some results and performances.
\end{abstract}

\section{Introduction}

Since the great success of the Standard Model in the recent decades, there has 
been no doubt that the gauge theories are capable to describe the interactions 
between elementary particles. With larger colliding energy available to probe 
higher energy scattering events, only remained piece of the Standard Model, 
Higgs boson(s), will be also discovered in the current and future colliders. 
Besides, a new physics might be opened in this high energy frontier. Among 
these energy scale, precise predictions by the perturbative calculations is 
crucial for the signal and background estimations because their event 
topologies become much complicate with increasing the colliding energy. We 
have carried on the automatic computation of the Feynman diagrams by 
GRACE \cite{grace} system since we immediately have a huge number of diagram 
calculation in multi-particle final state processes although we can, in 
principle, calculate them by hand based on their Lagrangian in perturbation 
theory.

GRACE has satisfied this requirement at one-loop level \cite{grcloop} in the 
electroweak interactions as well as at tree level and at the minimal 
supersymmetric extension of the Standard Model (MSSM) \cite{grcmssm}. Those 
development has been mostly aimed at applications to lepton collisions. The 
generated codes however are not directly applicable to hadron-collision 
interactions due to the presence of a parton distribution function (PDF). 
Also, a certain process in hadron collisions consists of lots of subprocesses 
by referring incoming partons in PDF or outgoing partons in jets. In current 
scheme, it leads much time for the diagram calculations. We clearly need an 
extended framework of the GRACE system for hadron collisions. Early extensions 
can be seen in \cite{abe} and \cite{odaka}.

In order to implement those features specific to hadron collisions, we have 
developed an extended framework, called GR@PPA (GRace At PP/Anti-p). The 
primary function of GR@PPA is to determine the initial and final state 
partons, $i.e.$ their flavors and momenta in the incoming partons by referring 
to a PDF and the final state parton configuration if the process requires jets 
or decay products from massive bosons. Based on the GRACE output codes, GR@PPA 
calculates the cross section and generates unweighted parton-level events 
using BASES/SPRING \cite{bases} included in the GRACE system. The GR@PPA 
framework also includes an interface for a common data format (LHA) 
\cite{leshouches} with the common interface routines proposed at Les Houches 
Workshop on 2001\cite{tevwork}. To make the events realistic, the unweighted 
event data are passed through the showering-MC of PYTHIA \cite{pythia} or 
HERWIG \cite{herwig} which implements the initial- and final-state radiation, 
hadronization and decays and so forth.

Although the GR@PPA framework is not process-specific and can be applied to 
any other processes in hadron collisions using the output codes of the GRACE 
system\footnote{The extension of GRACE itself to the hadron collisions is 
under development.}, we also provide some primitive processes packed as a set 
of matrix element customized for this extension. At the moment, the selected 
processes are boson(s) plus n jets processes and $t\bar{t}$ plus m jets 
processes, where n(m) is accounted for up to 4(1) jets. These processes are 
the most important background processes for the Higgs boson searches or the 
SUSY particle searches as well as the precise measurements for the 
understandings of multi-body particle dynamics. Our previous work, four bottom 
quark production processes (GR@PPA\_4b \cite{grappa_4b}), is also included. 
The reasons why we also provide the particular processes apart from the 
benefit of the automatic Feynman diagram calculation by the GRACE system are 
followings. First, kinematical singularities in each process are cared with a 
proper treatment. Since the kinematics are optimized to be suitable for 
well-convergence behavior, one can get high efficiency to generate unweighted 
events without any care. This is immediately addressed to the program running 
speed which is critical for the large scale MC production. Second, it is easy 
to adopt the higher order calculations. In higher order calculations, the 
calculation procedure is normally process-specific. To avoid 
the negative weight in cancellation between virtual and loop correction, one 
requires the phase space points to be positive differential cross section. 
This feature is so difficult to generalize by the automatic Feynman diagram 
calculation. Once the customized matrix element for the NLO process is 
prepared \cite{kuriharanlo}, one can simply use it. Third, some extensions are 
possible only in the modification of the framework. For example, using an 
ability of C++ language, the GR@PPA generators will work in the C++ 
environment just by rewriting the framework by C++ but the others are still 
Fortran. This is minimum changes to wrap the Fortran code produced by the 
GRACE system. The parton shower algorithm, for example NLL parton 
shower \cite{nll}, is also possible to implement by modifying the framework 
because the parton shower is not a process specific model. Note that the 
extended GRACE system for hadron collisions will work to provide a set of 
matrix elements apart from the GR@PPA framework at these point.

In this paper, we describe a symbolic treatment of the diagram calculation 
adopted in GR@PPA for hadron collisions in the next section. Some benchmark 
cross sections and program performances are presented in Section 3. Our 
numerical results were compared with several 
generators \cite{alpgen,madgraph,comphep,amegic}. We got a good agreement with 
them \cite{mc4lhc2003}. Finally, a summary is given in Section 4. 

\section{Extension of GRACE to $pp$/$p\bar{p}$ collisions}

In hadron-hadron collision, a certain process consists of several incoherent 
subprocesses according to colliding partons in the hadrons. If the given 
process has a decay or a jet in the final state, the whole possible 
combinations of the outgoing partons are also taken into account for. The 
total cross section is thus expressed as a simple summation of those 
subprocesses in 
Eq.(\ref{eq:xsec})
\begin{equation}
\sigma = \sum_{i, j, F} \int dx_{1} \int dx_{2} \int d\hat{\Phi}_{F} 
f^{1}_{i}(x_{1},Q^{2}) f^{2}_{j}(x_{2},Q^{2}) 
{ d\hat{\sigma}_{i j \rightarrow F}(\hat{s}) \over d\hat{\Phi}_{F} },
\label{eq:xsec}
\end{equation}
where $f^{a}_{i}(x_{a},Q^{2})$ is a PDF of the hadron $a$ ($p$ or $\bar{p}$), 
which gives the probability to find the parton $i$ with an energy fraction 
$x_{a}$ at a probing virtuality of $Q^{2}$. The differential cross section 
$d\hat{\sigma}_{i j \rightarrow F}(\hat{s})/d\hat{\Phi}_{F}$ describes the 
parton-level hard interaction producing the final-state $F$ from a collision 
of partons, $i$ and $j$, where $\hat{s}$ is the square of the total initial 
4-momentum. The sum is taken over all relevant combinations of $i$, $j$ and 
$F$. We had mainly two sorts of development in GR@PPA, --- applying PDF in the 
phase space integration and sharing several subprocesses as a single 
base-subprocess. The former is described in our previous 
paper \cite{grappa_4b}. Here, we focus on the later case. 

The original GRACE system assumes that both the initial and final states are 
well-defined. Hence, it can be applied to evaluating 
$d\hat{\sigma}_{i j \rightarrow F}(\hat{s})/d\hat{\Phi}_{F}$ and its 
integration over the final-state phase space $\hat{\Phi}_{F}$ only. An 
adequate extension is necessary to take into account the variation of the 
initial and final states both in parton species and their momenta, in order to 
make the GRACE system applicable to hadron collisions. As already mentioned, a 
"process" of interest is usually composed of several incoherent subprocesses 
in hadron interactions. However, in many cases, the difference between the 
subprocesses is the difference in the quark combination in the initial and/or 
final states only. The matrix element of these subprocesses is frequently 
identical, or the difference is only in a few coupling parameters and/or 
masses. In such cases, it is convenient to add one more 
integration/differentiation variable to replace the summation in 
Eq.(\ref{eq:xsec}) with an integration. As a result, these subprocesses can 
share an identical "GRACE output code" and can be treated as a single 
subprocess. This technique simplifies the program code and saves the computing 
time very much.

The number of the combinations taken N out of M flavors allowing to overlap
them, in general, is given by 
$_{M}H_{N}$($\equiv$ $\frac{(N+M-1)!}{N!(M-1)!}$). In case that all parton 
flavors are considered, 
$M=11(u,d,c,s,b,g,\bar{u},\bar{d},\bar{c},\bar{s},\bar{b})$, then the 
configuration of the N jets final state has $_{11}H_{N}$ subprocesses if we 
neglect the conservation by total amount of charges of this subprocess. 
Clearly we can see that smaller $M$ decreases the number of subprocesses. 
Unless the flavor difference is taken account in the process, the flavor 
configurations can be replaced by a generic up-type, down-type parton, and 
gluon ($_{5}H_{N}$ $\ll$ $_{11}H_{N}$). The base-subprocesses are thus 
configured only as to have those partons and gluon. The output code of the 
matrix element from the GRACE system is extended to have a function of input 
masses and couplings, so that the masses and couplings are interchanged 
according to the assigned flavors. Note that each base-subprocess covers every 
possible combination of flavors. The diagram selection in the 
base-subprocesses thus allows to specify all subprocesses with a proper flavor 
configuration. In addition, since the Feynman diagrams within the process of 
Standard Model are symmetry with respect to the momentum (parity) and charge 
flip of the initial colliding partons in the CM frame of the process, the 
subprocesses can be reduced more. In Table \ref{tab:subproc}, we list up the 
number of all possible base-subprocesses contributes in N jets process in 
$pp(\bar{p})$ collisions together with that of the subprocesses counted for 
all flavor combinations. The subprocesses are classified according to the 
difference in the initial-state parton combination. That is, the initial 
parton combinations in the base-subprocesses are 
$q_{u}q_{u}(\bar{q_{d}}\bar{q_{d}})$, $q_{u}\bar{q_{d}}$, 
$q_{u}g(\bar{q_{d}}g)$, $q_{u}q_{d}$, $q_{u}\bar{q_{u}}(q_{d}\bar{q_{d}})$, 
gg, where $q_{u}$($q_{d}$) and $g$ is up(down)-type quarks, and gluon, 
respectively. We take them all positive side.

The integration of Eq.(\ref{eq:xsec}) has a weight factor for each subprocess.
If a decay products from the resonance particles is separately taken account 
for the final state partons of N jets configuration, Eq.(\ref{eq:xsec2}) can 
be rewritten as 
\begin{equation}
\sigma = \int dw_{i,j,F} \int dx_{1} \int dx_{2} \int d\hat{\Phi}_{F} 
w_{i,j,F} \cdot \frac{d\hat{\sigma}_{i j \rightarrow F}^{selected}
(\hat{s};m,\alpha)}{d\hat{\Phi}_{F}} ,
\label{eq:xsec2}
\end{equation}
where $d\hat{\sigma}_{i j \rightarrow F}^{selected}$ is the differential cross 
section of each subprocess with an input arguments of the masses and 
couplings. The matrix element is supplied by that of the base-subprocess. 
Based on the initial and final state parton configuration, the graph selection 
is applied to this base-subprocess, and masses and couplings are given in the 
diagram calculation event by event. The $w_{i,j,F}$ is a weight factor for the 
initial and final state parton configuration, and can be expressed as 
\begin{equation}
w_{i,j,F} = \sum_{i,j,F} f^{1}_{i}(x_{1},Q^{2}) f^{2}_{j}(x_{2},Q^{2}) 
\cdot |V_{CKM}|^{2K} \cdot 
\{Br.(X \rightarrow F') \times \Gamma_{tot}^{X}\}^{L} ,
\label{eq:xsec3}
\end{equation}
where an index $K$ and $L$ is the number of W and $X$ bosons, respectively. A 
PDF is responsible for the weight of the initial state parton configuration, 
and a squared coupling normalized by that of the base-subprocess is 
responsible for the weight of the final state parton configuration. Note that 
the square of the CKM (Cabibbo-Kobayashi-Maskawa) \cite{ckm} matrix parameter 
remains after normalization of the coupling of the base-subprocess depending 
on the number of the W bosons $K$. If the $X$ boson presents in the Feynman 
diagrams and decay into $F'$ without interference with the other partons, 
where $F'$ is a member of the final state particles, then the fraction of the 
decay is used as the weight factor. The branching ratio and total width may be 
given by the experimental measured one.

\section{Results}

The total cross sections estimated by GR@PPA are presented in 
Table \ref{tab:xsecsgl} for the single boson plus jets productions and 
Table \ref{tab:xsecdbl} for the double bosons plus jets productions, 
respectively, where all bosons decay into electron and positron ($e^{+}e^{-}$) 
or electron(positron) and (anti-)electron-neutrino ($e\nu_{e}$). Both are 
shown with the cases of Tevatron Run-II ($p\bar{p}$ collisions at $\sqrt{s}$ 
$=$ 1.96 TeV) and LHC ($pp$ collisions at $\sqrt{s}$ $=$ 14 TeV) conditions 
with CTEQ5L \cite{cteq} PDF. The renormalization and factorization scales 
($Q^{2}$) are chosen to be identical, and those values are taken as the 
squared boson mass for the single boson productions and the summation of 
squared boson masses for the double bosons productions processes. The cuts are 
only applied for the final state partons (jets), with the values of 
$p_{T}$ $>$ 8.0 GeV, $|\eta|$ $<$ 3.0, and $\Delta$R $>$ 0.4 for Tevatron, and 
$p_{T}$ $>$ 20 GeV, $|\eta|$ $<$ 3.0, and $\Delta$R $>$ 0.4 for LHC 
conditions, but no cut is applied for leptons from bosons. The integration 
accuracy achieved in BASES is fairly better than 1\% for all processes with 
the default settings of the mapping number for the integrations.

The performance of GR@PPA for the W + N jets processes in Tevatron Run-II 
condition is also summarized in Table \ref{tab:xsecspeed}. Used processor is 
Intel Xeon 3.4 GHz processor. Tests are performed by two different Fortran 
compilers: a free software of g77 ver.2.96 and a commercial compiler of Intel 
Fortran Compiler ver.8.0. The integration time and the generation speed are 
separately shown. Clearly, the commercial compiler is $\sim$ 2.5 faster than 
the free compiler, but in both cases, those are not intolerable time for the 
large scale Monte Carlo production. The generation efficiencies by SPRING are 
also shown. These are within an order of a few percent for most of the 
processes. These numbers are exceptionally good for this kind of complicated 
processes. 

\section{Summary}

We have developed an extended framework, named GR@PPA, of the GRACE system for 
hadron collisions. We have introduced the scheme to share some subprocesses as 
a single subprocess. We found that this extension allows us to incorporate the 
variation in the initial and final states parton configurations into the GRACE 
system. The results for some processes with multi-parton configurations are 
presented, and we found that the computing time for the diagram calculations 
is drastically reduced to be compared with the original GRACE system which 
assumes that both the initial and final states are well-defined and the 
integration is performed for every each subprocess. Using this faculty of 
GR@PPA, we expect that the event generator is suitable not only for a large 
scale Monte Carlo production at high luminosity hadron colliders of Tevatron 
or LHC, but also for future NLO calculations which is also composed of lots 
of subprocesses.

\section{Acknowledgements}

The author would like to thank all the people of the Minami-Tateya numerical 
calculation group and the ATLAS-Japan Collaboration. The author would also 
like to thank the organizers of the Conference.




\begin{table}[htbp]
\begin{center}
\begin{tabular}{c|c|c} \hline
N jets ($\alpha_{s}^{N}$) & base-subproc. & suproc. w/ all flavors \\ \hline
2                         & 8             & 176                    \\
3                         & 9             & 276                    \\
4                         & 14            & 891                    \\ \hline
\end{tabular}
\end{center}
\caption{Number of all possible base-subprocesses contributes in N jets 
process in $pp(\bar{p})$ collisions together with one of the subprocesses 
counted for all flavor combinations. The subprocesses are classified according 
to the difference in the initial-state parton combination.}
\label{tab:subproc}
\end{table}

\begin{table}[htbp]
\begin{center}
\begin{tabular}{c|c|c|c|c|c|c} \hline
 & N jets          & 0       & 1        & 2        & 3       & 4     \\ \hline
Tevatron
 & $W(e\nu_{e})$ + & 2040(1) & 696.0(6) & 237.2(3) & 77.8(1) & 27.12(6) \\
($\sqrt{s}$ $=$ 1.96 TeV)
 & $Z(e^{+}e^{-})$ + & 1222(2) & 174.9(3) & 57.8(1) & 17.26(3) & 7.5(1) \\ \hline \hline
LHC
 & $W(e\nu_{e})$ + & 17892(12) & 3949(3) & 1607(2) & 682(2)  & 316(1) \\
($\sqrt{s}$ $=$ 14 TeV)
 & $Z(e^{+}e^{-})$ + & 7208(12) & 609(1) & 278.0(5) & 105.1(3) & 64.0(4) \\ \hline
\end{tabular}
\end{center}
\caption{The total cross section (pb) for the single boson plus jets 
productions estimated by GR@PPA, where all bosons decay into electron and 
positron ($e^{+}e^{-}$) or electron(positron) and (anti-)electron-neutrino
($e\nu_{e}$). Results are presented for the cases of Tevatron Run-II 
($p\bar{p}$ collisions at $\sqrt{s}$ $=$ 1.96 TeV) and LHC 
($pp$ collisions at $\sqrt{s}$ $=$ 14 TeV) with CTEQ5L. The renormalization 
and factorization scales are chosen to be identical, and those values are 
taken as the squared boson mass. The cuts are only applied for the final state 
partons (jets), with the values of $p_{T}$ $>$ 8.0 GeV, $|\eta|$ $<$ 3.0, and 
$\Delta$R $>$ 0.4 for Tevatron, and $p_{T}$ $>$ 20 GeV, $|\eta|$ $<$ 3.0, and 
$\Delta$R $>$ 0.4 for LHC conditions, but no cut is applied for leptons from 
the boson.}
\label{tab:xsecsgl}
\end{table}

\begin{table}[htbp]
\begin{center}
\begin{tabular}{c|c|c|c|c} \hline
 & N jets          & 0       & 1        & 2        \\ \hline
Tevatron
& $W^{+}(e^{+}\nu_{e})W^{-}(e^{-}\bar{\nu_{e}})$ + & 
109.1(2) & 60.31(9) & 31.45(9) \\
($\sqrt{s}$ $=$ 1.96 TeV)
& $Z(e^{+}e^{-})Z(e^{+}e^{-})$ + & 
4.932(7) & 2.295(3) & 0.895(1) \\
& $Z(e^{+}e^{-})W(e\nu_{e})$ + & 
17.85(3) & 10.40(2) & 4.61(1) \\ \hline \hline
LHC
& $W^{+}(e^{+}\nu_{e})W^{-}(e^{-}\bar{\nu_{e}})$ + & 
1425(4) & 906(2) & 2680(3) \\
($\sqrt{s}$ $=$ 14 TeV)
& $Z(e^{+}e^{-})Z(e^{+}e^{-})$ + & 
44.89(7) & 18.25(3) & 9.74(1) \\
& $Z(e^{+}e^{-})W(e\nu_{e})$ + & 
186.7(3) & 449.3(8) & 367(1) \\ \hline
\end{tabular}
\end{center}
\caption{The total cross section (fb) for the double bosons plus jets 
productions estimated by GR@PPA, where all bosons decay into electron and 
positron ($e^{+}e^{-}$) or electron(positron) and (anti-)electron-neutrino
($e\nu_{e}$). Results are presented for the cases of Tevatron Run-II 
($p\bar{p}$ collisions at $\sqrt{s}$ $=$ 1.96 TeV) and LHC 
($pp$ collisions at $\sqrt{s}$ $=$ 14 TeV) with CTEQ5L. The renormalization 
and factorization scales are chosen to be identical, and those values are 
taken as the summation of the squared boson masses. The cuts are only applied 
for the final state partons (jets), with the values of $p_{T}$ $>$ 8.0 GeV, 
$|\eta|$ $<$ 3.0, and $\Delta$R $>$ 0.4 for Tevatron, and $p_{T}$ $>$ 20 GeV, 
$|\eta|$ $<$ 3.0, and $\Delta$R $>$ 0.4 for LHC conditions, but no cut is 
applied for leptons from the boson.}
\label{tab:xsecdbl}
\end{table}

\begin{table}[b]
\begin{center}
\begin{tabular}{c|c|c|c|c} \hline
Process & Fortran  & Integration    & Event Generation & Efficiency \\
        & Compiler & time (H:M:Sec) & speed (events/sec) & (\%) \\ \hline
$W(e\nu_{e})$ + 0 jet & g77       & 0:0:4.88 &  43859 & 69.618 \\
                      & intel 8.0 & 0:0:1.80 & 101010 &        \\ \hline
$W(e\nu_{e})$ + 1 jet & g77       & 0:0:51.81 & 13927 & 19.891 \\
                      & intel 8.0 & 0:0:19.15 & 34364 &        \\ \hline
$W(e\nu_{e})$ + 2 jets & g77       & 0:38:26.66 &  708.5 & 1.731 \\
                       & intel 8.0 & 0:13:52.38 & 1956.5 &       \\ \hline
$W(e\nu_{e})$ + 3 jets & g77       & 14:03:46.35 & 17.99 & 0.283 \\
                       & intel 8.0 & 04:57:49.50 & 51.65 &       \\ \hline
\end{tabular}
\end{center}
\caption{Performance of GR@PPA for the W + N jets processes in Tevatron Run-II 
condition. Used processor is Intel Xeon 3.4 GHz processor. Tests are performed 
by two different Fortran compilers: a free software of g77 ver.2.96 and a 
commercial compiler of Intel Fortran Compiler ver.8.0. The integration time 
and the generation speed are separately shown. The generation efficiencies by 
SPRING are also shown.}
\label{tab:xsecspeed}
\end{table}

\end{document}